\documentstyle[preprint,aps]{revtex}

\begin{document}
\draft
\title{The Isovector Quadrupole-Quadrupole Interaction Used in Shell
Model Calculations}

\author{M.\ S.\ Fayache, S.\ Shelly Sharma\cite{A} and L.\ Zamick}
\address{Department of Physics and Astronomy, Rutgers University,
	Piscataway, New Jersey 08855}
\date{\today}
\maketitle

\begin{abstract}
An interaction $-\chi Q\cdot Q(1+B\vec{\tau}(1)\cdot \vec{\tau}(2))$
is used in a shell model calculation for $^{10}Be$. Whereas for $B=0$
the $2_1^+$ state is two-fold degenerate, introducing a negative $B$
causes an `isovector' $2^+$ state to come down to zero energy at
$B=-0.67$ and an $S=1~L=1$ triplet ($J=0^+,~1^+,~2^+$) to come down to
zero energy at $B=-0.73$. These are undesirable properties, but a
large negative $B$ is apparently needed to fit the energy of the
isovector giant quadrupole resonance.
\end{abstract}


\narrowtext

\section{Introduction}

In this work, we wish to deal with a mismatch which occurs when the
schematic quadrupole-quadrupole interaction, including an
isospin-dependent part, is used as a particle-hole interaction in
$R.P.A.$ calculations or is used as a particle-particle interaction in
shell model calculations. Concerning the former, we have the reviews
of  Bayman \cite{bayman} and of Bes and Sorensen \cite{bes} which show
that the $Q \cdot Q$ interaction plus pairing can explain the low
lying $2^+$ vibrational states in even-even nuclei. These are also
well described in the books of Bohr and Mottelson \cite{bm} and of
Soloviev \cite{solov}. Bes, Broglia and Nilsson \cite{bbn}, Bohr
and Mottelson \cite{bm} and Suzuki and Rowe \cite{suz} also use the $Q
\cdot Q$ interaction for high frequency modes and note that, in order
to explain the large splitting between
the isoscalar and isovector giant quadrupole resonances, one needs a
strong isospin-dependent term $Q \cdot Q \vec{\tau}(1) \cdot
\vec{\tau}(2)$. Using the parametrization $V=-\chi Q\cdot
Q(1+B\vec{\tau}(1)\cdot \vec{\tau}(2))$ (where
$\vec{\tau}(1) \cdot \vec{\tau}(2)=1$ for $T=1$ states and -3 for
$T=0$ states),
Bohr and Motteslon \cite{bm} state that $B$ is equal to -3.6, very
large indeed. Soloviev \cite{solov} quotes the formula
$B^{(\lambda)}=-0.5(2\lambda+3)$ which equals -3.5 for $\lambda=2$,
but in actual calculations he uses $-0.2(2\lambda+3)$. In Ref.
\cite{bbn} it is noted however that if $B=-3.6$ in a large space
including $\Delta N=2$ excitations, then if one truncates to a $\Delta
N=0$ space one should use a value which is much smaller in magnitude
$B=-0.6$. More recent references include those of Hamamoto and
Nazarewicz \cite{hz} and of Nojarov, Faessler and Dingfelder
\cite{nf8,nf9,nfd}. The latter authors made a critical study of the
parameter $B$ and concluded that it should have a smaller magnitude
than was previously used. They use $B=-2$ in Ref. \cite{nfd} then
compare this favoured value with other values e.g. $B=-3.6$ and
$B=-0.6$.

On the other hand, the $Q\cdot Q$ interaction has been used as a
particle-particle interaction as well, especially by Elliott with his
$SU(3)$ scheme \cite{ell}. In the $s-d$ shell this interaction is used to
explain rotational behaviour in many nuclei e.g. $^{20}Ne$, $^{22}Ne$
and $^{24}Mg$. The model, as shown by Elliott \cite{ell} and by
Harvey \cite{har}, also helps explain deviations from the extreme
rotational model due to the truncation effects in the shell model.
However, Elliott \cite {ell} uses a $Q\cdot Q$ interaction {\em
without} an isovector term (i.e. with $B=0$). One may well wonder what
would happen to his scheme if we introduced a large
$\vec{\tau}(1)\cdot \vec{\tau}(2)$ term.

In this work we consider precisely this problem but we  work
in the $p$ shell, where things are even simpler than in the $s-d$
shell, and consider the case of $^{10}Be$. We choose this nucleus
because it is strongly deformed ($\beta=1.12$ according to the tables
of Raman et. al. \cite{ram}) and also because it is an $N\not=Z$ nucleus. For
such a nucleus we can have isovector transitions from the ground state
which don't change the overall isospin. Such transitions are very
important for our considerations.

\section{Calculations}

We perform $p$ shell calculations for states in $^{10}\mbox{Be}$ using
the interaction

\begin{equation}
V=-\chi Q\cdot Q(1+B\vec{\tau}(1)\cdot \vec{\tau}(2))
\end{equation}

\noindent with $\chi=0.36146$. We study the behaviour of selected
states for various negative values of $B$, the coefficient of the
isovector $Q\cdot Q$ interaction.

An attractive feature of the multipole-multipole interaction
$O^{L\tau} \cdot O^{L\tau}$ is that direct particle-hole matrix
elements of the form $\langle
(P_BH_B^{-1})^{JT}~V~(P_AH_A^{-1})^{JT}\rangle$ vanish unless $J=L$
and $T=\tau$ (the exchange terms are usually taken to be zero). Thus
if one adjusts the parameters of one $(JT)$ mode pictured as an
$R.P.A.$ state $\Psi^{JT}=\sum
\{X_{PH}(PH^{-1})^{JT}~-~Y_{PH}(PH^{-1})^{\dagger ~JT}\}$, then the other
modes $(JT)'\not=(JT)$ are not affected.

The expressions that we use for the particle-particle matrix elements
are as follows:

\begin{eqnarray*}
\left\langle[j_A(1)j_B(2)]^{JT}~|-\chi Q\cdot
Q[(1+B\vec{\tau}(1)\cdot
\vec{\tau}(2))]~|~\{[j_C(1)j_D(2)]^{JT}-[j_C(2)j_D(1)]^{JT}\}\right\rangle~=\\
-\chi
[1+(\delta_{T,1}-3\delta_{T,0})B][1+(-1)^{j_C+j_D+J+T}P_{j_Cj_D}]
\left ( \frac{2j_A+1}{2j_C+1}\right )^{\frac{1}{2}}\\
\times \langle \Psi^{j_A}[Y^Lr^L\Psi^{j_C}]^{j_A}\rangle
\langle \Psi^{j_B}[Y^Lr^L\Psi^{j_D}]^{j_B}\rangle
U(\begin{array}{llllll}j_A & L & J & j_D; & j_C & j_B \end{array})
\end{eqnarray*}

\noindent where we define our singly reduced matrix elements by the
following convention:

\[(L~j_1~M~m_1|j_2~m_2)\langle \Psi^{J_2}[O^L\Psi^{J_1}]^{J_2}\rangle=
\langle \Psi^{J_2}_{M_2}~O^L_M~\Psi^{J_1}_{M_1}\rangle \]

\noindent In the above $U$ is the unitary Racah coefficient. Note that
the entire dependence on $J$, the total angular momentum of the two
particles, is contained in the above $U$ coefficient.

The expression for the {\em direct} part of the particle-hole
interaction is:

\begin{eqnarray*}
\langle (j_Aj_B^{-1})^{JT}~|V|~(j_Cj_D^{-1})^{JT}\rangle=-\chi
(\delta_{T,0}+B\delta_{T,1})[4(2j_A+1)(2j_C+1)]^{\frac{1}{2}}\delta_{J,L}\\
\times \langle \Psi^{j_B}[Y^Lr^L\Psi^{j_A}]^{j_B}\rangle
\langle \Psi^{j_C}[Y^Lr^L\Psi^{j_D}]^{j_C}\rangle
\end{eqnarray*}

In the above expression we have made the isospin dependence as
explicit as possible. We see that for particle-hole states the
$T=0$ shift is proportional to $-\chi$ and the $T=1$ to $-\chi B$. We
can give a simplified derivation of the value of $B$ using the
quadrupole giant resonance data presented in Suzuki and Rowe's work
\cite{suz}. The isoscalar quadrupole resonance is at
$\frac{63}{A^{\frac{1}{3}}}~MeV$. The unperturbed energy of these
giant resonances is $2\hbar \omega=\frac{82}{A^{\frac{1}{3}}}~MeV$.
Assuming a Tamm-Dancoff model, the ratio of isovector to isoscalar
shifts is

\[\frac{-\chi B}{-\chi}=B=\frac{141-82}{63-82}=-3.1\]

\subsection{The $B=0$ Limit}

We now consider shell model calculations for $^{10}Be$ using the
interaction of Eq.(1).
For $B=0$ we simply have the interaction $-\chi Q\cdot Q$. In the $p$
shell, as for any spin-isospin independent interaction, the states are
classified by the orbital symmetry values $[f_1~f_2~f_3]$ as well as
by the quantum numbers $L,~S$ and $T$ \cite{ham}. The energies are
given by the
Elliott formula \cite{ell} for $SU(3)$ with $\lambda=f_1-f_2$ and
$\mu=f_2-f_3$:

\begin{equation}
E=\bar{\chi}[-4(\lambda^2+ \mu^2+\lambda \mu)+3(\lambda+\mu)+3L(L+1)]
\end{equation}

where $\bar{\chi}=\frac{5b^4}{32\pi}\chi$ and $b$ is the harmonic
oscillator length parameter ($b^2=\frac{\hbar}{m\omega}$). The values
of $\chi$ and $\bar{\chi}$ for $^{10}Be$ are respectively 0.36146 and
0.1286. The ground state has quantum numbers $[4~2~0]$ $S=0~L=0~T=1$;
$J=0^+$. The first excited state is doubly degenerate: $L=2~S=0$
$[4~2~0]$; $J=2_1^+,~2_2^+$, and the excitation energy is
$18\bar{\chi}$. We  also consider the next excited states
arising from
two degenerate orbital symmetry states $[3~3~0]$ and $[4~1~1]$. The
other quantum numbers are $L=1~S=1~T=1$. The $J$ values are therefore
$0^+$, $1^+$ and $2^+$. In other words, for each orbital symmetry we
have a triplet of states. The excitation energy is $30\bar{\chi}$.
These states cannot be reached from the ground state by the $M1$
operator. There are several other states, one of which is the
scissors mode state with quantum numbers $L=1~S=0$ ($J=1^+$). This
state has an excitation energy of $66\bar{\chi}$ and orbital symmetry
$[f]=[3~2~1]$. Of particular interest is the fact that the $T=1$ and
$T=2$ scissors mode states are degenerate in energy for the above
interaction $-\chi Q\cdot Q$. Note that the scissors mode is not the
lowest $1^+$ state; the aforementioned $L=1~S=1$ $J=1^+$ states lie
lower.

\subsection{The Dependence on $B$}

The main thrust of the paper is in this section. Having noted in the
introduction that a large and negative value of $B$ is needed to fit
the splitting of isovector and isoscalar giant quadrupole resonances,
we will now study what happens to selected states in $^{10}Be$ when a
finite negative $B$ is introduced. The results are presented in Fig.
1.

We first focus on the two $J=2^+$ states, which for $B=0$ are
degenerate and lie lowest. This is not the case experimentally; the
$2^+_1$ state is at 3.368 $MeV$ and the $2^+_2$ state at 5.960 $MeV$
the splitting being largely due to a spin-orbit interaction.
Both $2^+$ states have the same orbital
symmetry as the $J=0^+$ ground state $[f]=[4~2~0]$. We see from Fig.1
that as $B$ is made negative the degeneracy is removed with one state
going rapidly down towards zero energy and the other rising in energy.
The behaviour is not precisely linear, but the linear approximation
(which would hold if there were no admixtures of states of different
symmetry) is remarkably good for negative $B$. A linear fit to the
behaviour for negative $B$ is as follows:

\[\frac{E_A}{\bar{\chi}}=18-26.935|B|\]

\noindent with $E_A$ vanishing at $B=-0.668$.

The fact that the $2_A^+$ state comes down towards zero already is a
signal of a very peculiar behaviour. We find even more peculiar
behaviour if we look at the transition rates in Table $I$. For electric
quadrupole excitations we define $B(E2,e_p,e_n)$ as the transition
rate when effective charges $e_p$ and $e_n$ are used for the proton
and neutron respectively.

In Table $I$ we list the isoscalar transition rate $B(E2,1,1)$ and the
isovector transition rate $B(E2,1,-1)$. Note that the state
$|2_A\rangle$ comes down in energy as $B$ becomes more negative and is
excited from the ground state only by the isovector operator, whilst
the state $|2^+_B\rangle$ goes up in energy as $B$ becomes more negative
and is excited only by the isoscalar operator. This
behaviour, which is shown in Fig.1, clearly goes against experiment.
The lowest $2^+$ states in essentially all nuclei, although they may
have some isovector part, are dominantly isoscalar.

We now look at other selected states. The states which for $B=0$ have
quantum numbers $S=1~L=1$ $f=[3~3]$ and $[f]=[4~1~1]$ (two degenerate
configurations) lead to two sets of triplets with total angular
momenta $J=0^+,~1^+$ and $2^+$. When a finite negative $B$ is turned
on, the $J$ degeneracy is maintained but the degeneracy between the
two sets of triplets is removed. Both sets come down in energy, and in
the linear approximation we get:

\[\frac{E_C}{\bar{\chi}}=30-41.160|B|\]

\noindent with $E_C$ vanishing at $B=-0.729$, and

\[\frac{E_D}{\bar{\chi}}=30-31.850|B|\]

\noindent with $E_D$ vanishing at $B=-0.942$. We show the behaviour of
the state $|C\rangle$ as a function of $B$ in Fig.1. This figure shows
clearly the linear collapse of this state as well as the isovector
$2^+$ state $|A\rangle$ as a function of negative $B$.

Note that the $J=0^+,~1^+,~2^+$ triplet $|C\rangle$ vanishes at a
value of $B$ very close to that for the $J=2^+$ state $|A\rangle$.
The values are $B=-0.729$ and $B=-0.668$ respectively. Thus, care must
be taken not to confuse the $2^+$ states of each configuration in this
region of $B$ and beyond. There is a small region of $B$ from -0.67 to
-0.84 where the state $J=2_A^+$ is the lowest state. But then the
triplet $J=0_C^+,~1_C^+,~2_C^+$, although starting from a higher
energy at $B=0$, has a slope of larger magnitude than the $2_A^+$
state, and ultimately becomes the ground state for $B\leq-0.84$.

We finally look at the states with orbital symmetry $[4~2~1]$. One of
these states is the $L=1~S=0~T=1$ scissors mode state and is therefore of
special interest \cite{bo}, \cite{ia}. Equally of interest is the
other part of the scissors mode strength $L=1~S=0~T=2$. The behaviour
as a function of $B$ is shown in Fig.2 and Table $II$. For $B=0$ the
$T=1$ and $T=2$ scissors are degenerate in energy \cite{fay,ham}, and
there are four degenerate states in all for each $T$. As $B$ is made
negative two $T=1$ states come down in energy and two come up. We give
formulae for only the two extreme states -one going down the fastest
and one going up the fastest (respectively):

\[\frac{E_E}{\bar{\chi}}=66 -48.362|B|\]

\[\frac{E_F}{\bar{\chi}}=66+23.761|B|\]

By looking at the $M1$ rates at Table $II$, we see that the state
$J=1^+_F~T=1$ and $J=1^+_G~T=2$ form the scissors modes  -they get
{\em all} the isovector orbital strength. The state $|D\rangle$ which
goes down in energy has no isovector orbital strength. In Fig.2 we
show the behaviour, as a function of $B$, of the $T=1$ and $T=2$
scissors modes. Note that whereas for $B=0$ the two are degenerate,
for negative $B$ the $T=2$ strength comes below the $T=1$ strength,
another peculiar result.

We show now in Table $II$ the isoscalar and isovector (scissors mode)
orbital magnetic dipole rates. The transitions are to one state. From
$B=0$ to $B=-0.6$ the isovector rate to $T=1$ final states increases
from $0.0890~\mu_N^2$ to $0.230~\mu_N^2$ i.e. an isovector $Q\cdot Q$
interaction with negative $B$ causes the scissors mode strength to
increase. Conversly, for positive $B$, the strength decreases with
increasing $B$. It should be noted that for $B=0$ the strength is
$\frac{9}{32\pi}=0.0890\mu_N^2$. It should also be noted that in this
limit the $T=2$ scissors mode is degenerate in energy with the $T=1$
mode, and the strength to $T=2$ is $\frac{15}{32\pi}~\mu_N^2$. The
ratio is $\frac{(2T+1)_{T=2}}{(2T+1)_{T=1}}=\frac{5}{3}$. The
isoscalar orbital rate starts at {\em zero} for $B=0$ and increases
with negative $B$ to a finite albeit very small value i.e. for
$B=-0.6$ $B(M1)\uparrow_{isoscalar~orbital}=2.52\times10^{-3}~\mu_N^2$.

The energy-weighted $M1$ orbital strength also increases as $B$ is
varied. The combined $T=1$ and $T=2$ energy-weighted strength in the
range $0>B>-0.6$ is given by the approximate linear formula

\[EWS(B)=EWS(B=0)(1~-~1.7B)\].

\section{Beyond the Cross Over Region}

When $B$ becomes less than $-0.67$, the state $|A\rangle$ which is a
$J=2^+$ `isovector state' becomes the ground state. It so remains in the
range $-0.67>B>-0.84$. For $B<-0.84$ the ground state becomes
a triplet $S=1~L=1$ $J=0^+,~1^+,~2^+$emanating from some combination
of the states of orbital symmetry $[3~3~0]$ and $4~1~1]$ (these
orbital states are degenerate in energy at $B=0$).For $B$ sufficiently
negative, the nature of the ground state will again change.

But let us focus on the region of $B$ for which the triplet above is
the ground state ($B<-0.84$). Besides the striking fact that the
ground state {\em is} a triplet, what other evidence do we have of a
`phase transition' relative to the case where the orbital symmetry was
$[4~2~0]$? Let us consider the case $B=-1.0$ and look again at tables
$I$ and $II$ below the horizontal double lines.

 From table $III$ we see that the main {\em isovector} $E2$ strength is from
the $J=0^+$ to the $J=2^+$ member of the ground state triplet i.e. a
zero-energy transition. Although the $B(E2)$ is substantial,
$-6.80~e^2fm^4$, the rate would be zero if it is indeed a zero-energy
transition. The isoscalar $E2$ strength is now split almost evenly
between a low-energy state at $0.97~MeV$ and a high-energy state at
$8.7~MeV$. This is quite different from the case $B>-0.67$, where
all the strength was concentrated in one state.

Most interestingly a rather large isoscalar
orbital $M1$ strength emerges ($0.119~\mu_N^2$). It is again a `zero
energy' transition, however from $L=1~S=1~J=0^+$ to $L=1~S=1~J=1^+$.
Recall that for $B=0$ the isoscalar orbital strength is zero because
the ground state has $L=0$.
The isovector scissors mode strength from $J=0^+~T=1$ to the
$J=1^+~T=1$ states is now fragmented into two parts -a low-energy part
at $0.97~MeV$ with $B(M1)_{orbital}=0.13~\mu_N^2$ and a high-energy
part at $13.3~MeV$ with $B(M1)_{orbital}=0.27~\mu_N^2$. For
$B>-0.67$ all the $1^+,~T=1$ strength went only to one state.

\section{Closing Remarks}

We have shown that an isovector quadrupole-quadrupole interaction
$-\chi B Q(1)\cdot Q(2)\vec{\tau}(1)\cdot \vec{\tau}(2)$ with a large
negative $B$ yields very undesirable properties in $p$ shell model
calculations of $^{10}Be$. On the other hand, such interactions appear
to be needed to give correct splittings of the isovector and isoscalar
giant quadrupole resonances. From our analysis of $^{10}Be$ in a $p$
shell calculation, it appears that the best fit is obtained with $B=0$
(Elliott $SU(3)$ model) -even a small {\em positive} $B$ might be
acceptable.

For $B\simeq-0.7$ we get two sets of states collapsing to zero energy
-first an isovector $J=2^+$ state, and then an $L=1~S=1$ triplet
$J=0^+,~1^+,~2^+$. This behaviour is undesirable -no known nuclei
behave in this way. Perhaps a remedy to this dilemma is to introduce
momentum-dependent quadrupole terms in the interaction as Elliott had
done \cite{ell}.
This enabled him to have an interaction which did not connect the
$\Delta N=0$ with the $\Delta N=2$ space. We would choose these
so that the $\Delta N=0$ isovector quadrupole interaction is much
weaker than the $\Delta N=2$ part.

Another suggestion is to bring effective mass into the picture when
analyzing the separation of the isoscalar and isovector giant
quadrupole resonances. The unperturbed energy, rather than being $2
\hbar \omega$, is now $\frac{2 \hbar \omega}{m^*}$. The energy of the
isoscalar quadrupole resonance, rather than being $\sqrt{2}\hbar
\omega$ is now
$\sqrt{2}\frac{\hbar \omega}{(\frac{m^*}{m})^{\frac{1}{2}}}$
\cite{bm,goz,bert,bl}. Using the `empirical'
energies $\frac{63}{A^{\frac{1}{3}}}$ and
$\frac{141}{A^{\frac{1}{3}}}$ for the isoscalar and isovector
quadrupole resonances respectively, we now modify the estimate of $B$
in Sec. 2 as

\[B=\frac{141-\frac{82}{m^*}}{63-\frac{82}{m^*}}\]

\noindent With $\frac{m^*}{m}=0.8$ we get $B=-0.97$; with
$\frac{m^*}{m}=0.7$ we get $B=-0.41$. These are much smaller in
magnitude than the value $B=-3.1$ we obtained with $\frac{m^*}{m}=1$.
Furthermore, in a $\Delta N=0$ space, the magnitude of $B$ will be
even smaller due to renormalization effects \cite{bbn}.
This argument is admittedly somewhat hybrid, but we believe it
corresponds more closely to what happens when realistic interactions
are used.

\section{Acknowledgement}

This work was supported by the Department of Energy Grant No.
DE-FG05-40299. S.S. Sharma would like to thank the Physics Department
of Rutgers University for its hospitality and to acknowledge financial
support form $CNPq$, Brazil. We thank I. Hamamoto, R. Nojarov and
J. Millener for useful communications.

\squeezetable
\begin{table}
\caption{The energies and $B(E2,e_p,e_n)\uparrow$ (in $e^2fm^4$) to
the first two $2^+$ states as a function of $B$}
\begin{tabular}{ccccc}
  & \multicolumn{2}{c}{\underline {$2^+_{A}$(isovector)}}\tablenotemark[1] &
\multicolumn{2}{c}{\underline {$2^+_{B}$(isoscalar)}}\tablenotemark[1] \\
B & E (MeV) & $B(E2,1,-1)$ & E (MeV) & $B(E2,1,1)$\\
\tableline
0.6 & 3.49 & 42.04 & 4.02 & 57.85 \\
0.4 & 3.41 & 39.02 & 3.03 & 63.75 \\
0.2 & 3.00 & 35.03 & 2.48 & 67.33 \\
\tableline
 0.0 & 2.32 & 33.40 & 2.32 & 65.09 \\
\tableline
-0.2 & 1.57 & 30.26 & 2.44 & 67.71 \\
-0.4 & 0.87 & 29.49 & 2.77 & 66.45 \\
-0.6 & 0.21 & 29.11 & 3.21 & 65.06 \\
\end{tabular}
\tablenotetext[1] {The value of $B(E2,1,1)$ to the state
$|2_A^+\rangle$ is zero. The value of $B(E2,1,-1)$ to the state
$|2_B^+\rangle$ is zero.}
\end{table}

\begin{table}
\caption{The magnetic dipole {\em orbital} strengths (in $\mu_N^2$)
-both isoscalar and isovector- as a function of $B$}
\begin{tabular}{ccccccc}
 & \multicolumn{2}{c}{\underline {ISOSCALAR}} &
\multicolumn{4}{c}{\underline {ISOVECTOR}} \\
 & E & $1^+,~T=1$ & E & $1^+,~T=1 (scissors)$ & E & $1^+,~T=2$ \\
 & (MeV) & $B(M1)\uparrow_{orbital}$ & (MeV) &
$B(M1)\uparrow_{orbital}$ & (MeV) & $B(M1)\uparrow_{orbital}$\\
0.6 & 6.04 & 0 & 9.27 & 0.0040 & 11.74 & 0.0390 \\
0.4 & 5.55 & 0 & 8.59 & 0.0043 & 10.23 & 0.0728 \\
0.2 & 4.89 & 0 & 8.34 & 0.0370 & 9.16 & 0.1012 \\
\tableline
0.0 & 3.86 & 0 & 8.49 & 0.0895 ($\frac{9}{32\pi})$ & 8.49 & 0.1492
($\frac{15}{32\pi}$) \\
\tableline
-0.2 & 2.72 & 0 & 8.93 & 0.1431 & 8.10 & 0.1776 \\
-0.4 & 1.64 & 0 & 9.56 & 0.1922 & 7.91 & 0.2019 \\
-0.6 & 0.62 & 0 & 10.31 & 0.2314 & 7.84 & 0.2284 \\
\end{tabular}
\end{table}

\begin{table}
\caption{Beyond the Cross Over Region, $B=-1.0$}

\begin{tabular}{ccc}
$J=2^+,~T=1$  &  & \\
 $E^*(MeV)$ & $B(E2,1,1)$ ($e^2fm^4$)& $B(E2,1,-1)$ ($e^2fm^4$)\\
\tableline
0.00 & 0 & 6.7120 \\
0.96  & 15.93 & 0\\
8.68  & 21.68 & 0\\
\tableline
$J=1^+,~T=1$  &  & \\
 $E^*(MeV)$ & $B(M1)_{orbital}$ ISOSCALAR  ($\mu_N^2$)&
$B(M1)_{orbital}$ ISOVECTOR ($\mu_N^2$)\\
0.00 & 0.1194 & 0  \\
0.96  & 0 & 0.1368\\
13.3  & 0 & 0.2672\\
\end{tabular}
\end{table}

\pagebreak

{\large{\bf {Figure Captions}}}
\vspace{0.5in}

{\bf Figure (1):} The excitation energies of selected states in
$^{10}Be$ as a function of the isovector quadrupole interaction
parameter $B$. The solid line is for the $2_A^+$ state (isovector
$2^+$), the dashed line is for the $2_B^+$ state (isoscalar $2^+$)
and the dot-dashed for the $S=1~L=1$ triplet
($J=0^+_C,~1^+_C,~2^+_C$).

{\bf Figure (2):} Same as Figure 1 but for the $J=1^+~T=1$ scissors
mode branch (solid line) and for the $J=1^+~T=2$ scissors mode branch
(dashed line).

\end{document}